\documentclass[a4paper]{jpconf}
\usepackage{graphicx}
\begin{document}
\title{Systematic effects on the diversity of dwarf galaxies rotation curves}

\author{Juan M. Pacheco-Arias$^{1}$, Cristian R. Carvajal-Bohorquez$^{1}$, Juan C. B. Pineda$^{1}$ and Luis A. Núñez$^{1,2}$}

\address{$^{1}$ Escuela de F\'isica, Universidad Industrial de Santander, Carrera 27 Calle 9, Bucaramanga, Colombia.}

\address{$^{2}$ Departamento de F\'isica, Universidad de Los Andes, Mérida 5101, Mérida, Venezuela.}

\ead{juan.pacheco@correo.uis.edu.co}

\begin{abstract}
Cosmological simulations of structure formation are invaluable to study the evolution of the Universe and the development of galaxies in it successfully reproducing many observations in the context of the cosmological paradigm $\Lambda$CDM. However, there are remarkable discrepancies with observations that are a matter of debate. One of the most recently reported is the diversity of shapes in the rotation curves of dwarf galaxies in the local Universe which is in contrast to the apparent homogeneity of rotation curves in cosmological hydrodynamic simulations. Previous studies on similar problems have shown that sometimes can be alleviated by accounting for the impact of observational effects in the comparison. For this reason, in this work we present a set of controlled experiments to measure the impact that some systematic effects, associated with modeling the observation process in a realistic way, have on the diversity of synthetic rotation curves. Our results demonstrate that factors such as spectral power, spatial resolution and inclination angle, can naturally induce noticeable fluctuations on the shape of the rotation curves, reproducing up to $47\%$ of the diversity reported in the observations. This is remarkable, especially considering that we limited the sample to highly-symmetric disks simulated in isolation. This shows that a more realistic modeling of synthetic rotation curves may alleviate the reported tension between simulations and observations, without posing a challenge to the standard cosmological model of cold dark matter.
\end{abstract}

\section{Introduction}

Galaxies are the fundamental pieces that make up the large-scale structure of the Universe. They are an active subject of research, and provide substantial information for understanding the history of the cosmos [1]. Many catalogs of observations have been built overtime to try to understand the processes driving galaxy formation and evolution. These consistent efforts have established general trends in the characteristics of these objects [2], but they have also shown that there is a great diversity in the morphological, dynamical and spectral properties of galaxies. Understanding the origin of this diversity is one of the most important open questions in extragalactic astrophysics [3].

In this context, one of the methods that have gained more relevance in the study of the evolution of the Universe is the use of hydrodynamic numerical simulations, based mainly on the standard $\Lambda$CDM cosmological model. These simulations have successfully replicated the formation of large-scale structures in the Universe and, more recently, have been able to assemble realistic populations of galaxies consistent with multiple observations [4]. However, remarkable discrepancies have also been reported.

Two of the best-known discrepancies between observations and cosmological simulations that are in the debate are the cusp-core problem [5,6] and the unexpected diversity of dwarf galaxy rotation curves in the local Universe [7]. These issues are of great interest as they cast doubts on the ability of simulations to reproduce the range of properties observed in real galaxies, and raise questions about the standard cosmological model itself.

In the case of the cusp-core problem, it has been shown that the action of systematic effects related to the observation process may generate the illusion of detecting core-type dark matter density profiles in cuspy halos. This hypothesis is supported by investigations using controlled experiments with simulations, showing that observational factors affecting rotation curves such as symmetry deviations, pressure support, non-circular motions, spatial resolution, distance, and inclination angle, may easily generate such an illusion [8].

Following this line, the aim of this work is to identify if the inclusion of possible observational systematic effects, increases diversity in simulated rotation curves. To do so, we used realistic mock observations generated from galaxy simulations, which allowed us to establish the increase in the diversity of the rotation curve shapes once some of the most important limitations affecting observations are properly modeled during the analysis.

\section{Mock observations}

In this work we generated mock observations from six isolated dwarf galaxies simulated with a modified version of \textsc{gadget}-2 [9]. Simulated dwarf galaxies are ideal disks in rotational equilibrium
modeled after the main scaling relations for disk galaxies in the local Universe. These systems were presented in [8]. Each galaxy was analyzed in three different moments.

We model two different types of observations, including long-slit rotation curves and 2D velocity maps. We consider two different types of emission, labeled as $H_I$ and $H_{\alpha}$. We assume that each gas particle emits radiation through a spectrum composed of a single emission line with a Gaussian profile, discretized to create an IFU type data cube after assuming an orientation of the galaxy in the sky following a methodology similar to that in [8]. Yet in this work we use a more physical approach to estimate the amount of $H_{\alpha}$ radiation coming out of the gas particles, taking into account the ionization state of the gas to estimate the rate of recombinations. The instrumental parameters for the generation of the synthetic data cube, were set with the same values reported in the literature for the 21-cm and $H_{\alpha}$ local Universe observations. Each snapshot was {\it observed} with five inclination angles: $15^{\circ}$, $30^{\circ}$, $45^{\circ}$, $60^{\circ}$ and $75^{\circ}$, and with three different values for the resolution: $0.2$, $0.4$ and $0.8$ kpc. These values were selected to reproduce the characteristics exhibits in the catalog presented in the unexpected diversity problem [7].

As a result of the systematic variation of these parameters over the sample of simulated galaxies, a total of 2250 synthetic data cubes were generated and analyzed in the same way as it is done with observations, in order to extract the synthetic rotation curves which could then be compared with the theoretical curves drawn directly from the dynamic information of the particles in the simulation.

\section{Results}

To measure the variability we will follow exactly the same proposal of Oman, against whose results we want to compare, and this consists of measuring the circular velocity at 2 kpc of the radius ($V_{2kpc}$) and the maximum circular velocity of the rotation curve ($V_{max}$) to be contrasted. In that way, the variance in $V_{2kpc}$ for all curves with simulated values of $V_{max}$, encloses a measure of the diversity of curve shapes in a single parameter. Rotation curves obtained directly from gas particles in the simulations, without considering observational effects, are self-similar and have low variability in their central kiloparsecs, contrary to what is shown by observations (see figure 1), exactly as pointed out by [7]. For similar values of maximum circular speed, the circular speed at 2 kpc exhibits a low variability, reaching just between 2.1\% and 8.6\% of the variability exhibited by observations.

On the other hand, when the observational and instrumental factors are taken into account in the modeling of the rotation curves of the simulated galaxies, the scenario for the variability in the central region of the curves changes drastically (see figure 2). When all the mock rotation curves are taken into account, for the different combinations of observational parameters, the variability of the circular velocity at 2 kpc represents as much as 47\% of the observed variability.

Finally, when we model the effect of errors such as including noise, errors in distance estimation, or tilts in the angle of the slit, as commonly reported in the literature for these kinds of observations, the reported variability increases by 20\%, thus reproducing up to 67\% of the variability reported in the observations. Considering that in this study we only use highly asymmetric galaxies, with a very low amount of non-circular motions, it is very likely that this number gets higher if a sample of more realistic galaxies from simulations were used, i.e., galaxies formed in a fully cosmological context, experiencing cosmological torques and non-circular motions induced by mergers and close encounters with other galaxies, for instance. Non-circular motions must be taken into account when considering the diversity reported in the rotation curves for the sample of galaxies analyzed in this study [10].

\begin{figure}[h]
\begin{minipage}{14pc}
\includegraphics[width=21pc]{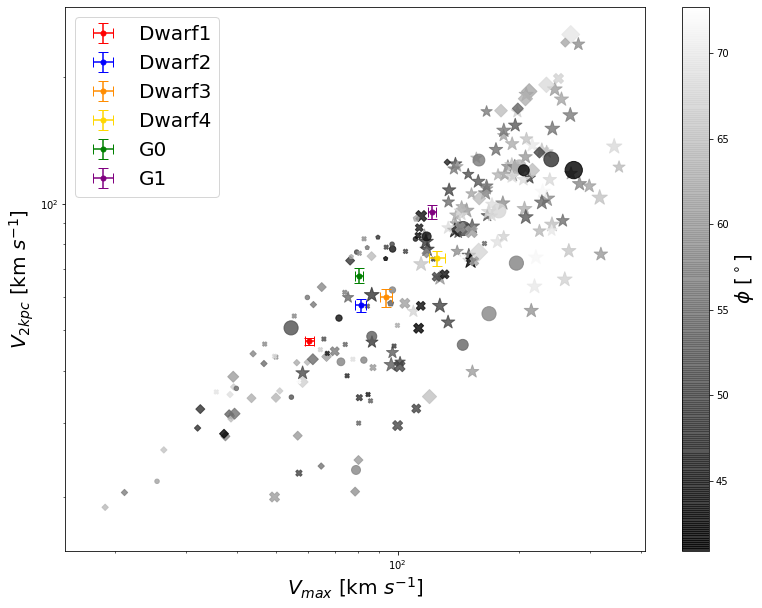}
\caption{\label{label} Circular speed at 2 kpc Vs maximum circular speed of the rotation curve. Grayscale points represent the observed galaxies reported in [7], while the colored ones identify the 6 simulated galaxies used for our experiments. Here we consider only the rotation curves created in an ideal theoretical way, without modeling the observational effects. Error bars indicate the standard deviation in each value for the whole sample of snapshots used, thus measuring the intrinsic variability of the theoretical circular velocity profiles, traditionally used to describe the dynamics of galaxies in simulation studies.}
\end{minipage}\hspace{6pc}%
\begin{minipage}{14pc}
\includegraphics[width=21pc]{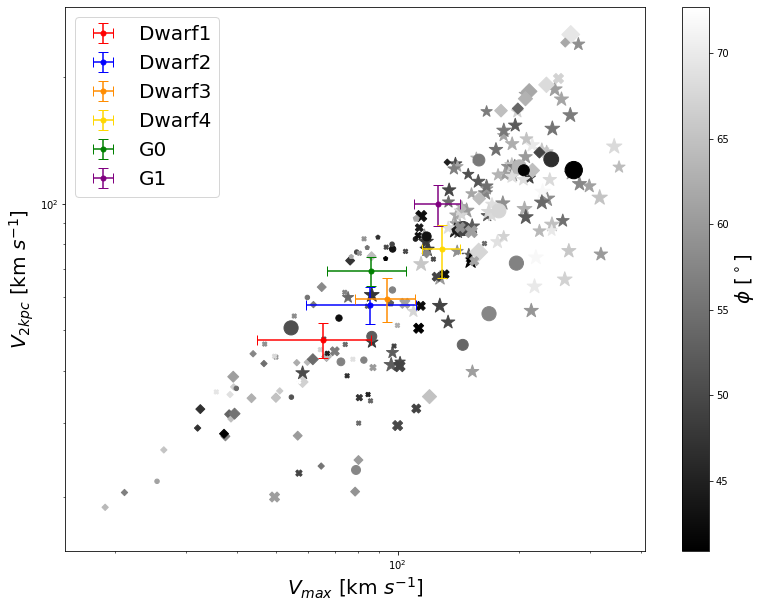}
\caption{\label{label} Circular speed at 2 kpc Vs maximum circular speed of the rotation curve, this time modeling the quoted observational effects to generate mock realistic rotation curves out from the simulations. The colors and symbols are the same as in figure 1. Error bars indicate the standard deviation in each value; it is measured across the total number of synthetic rotation curves, thus represents the variability achieved when a more realistic scenario is put into place, modeling the observation process in a  realistic manner.}
\end{minipage} 
\end{figure}

\section{Conclusion}

Realistic modeling of observational factors in the synthetic observation process increases the diversity in the rotation curves for the simulated galaxies. Combinations of values for these parameters can generate up to 47 \% of the variability reported in [7]. This allows alleviating the reported tension between observations and simulations, in this regard without resorting to exotic solutions like changing the baryonic physics or the cosmological model itself. We find that the discrepancy between the observed shapes of dwarf galaxies rotation curves may be originated in the way that data is used to perform the comparisons, thus not representing a real concern for the current status of hydrodynamic cosmological simulations.

\end{document}